\documentclass[12pt]{article}
\def\beq{\begin{equation}}
\def\eeq{\end{equation}}
\def\bea{\begin{eqnarray}}
\def\eea{\end{eqnarray}}
\def\beas{\begin{eqnarray*}}
\def\eeas{\end{eqnarray*}}
\def\nn{\nonumber}
\def\ra{\rangle}

\def\lb{[\![}
\def\rb{]\!]}
\def\t{\theta}

\begin{document}
\addtolength{\baselineskip}{2mm}
\addtolength{\abovedisplayskip}{1mm}
\addtolength{\belowdisplayskip}{1mm}
\addtolength{\parskip}{2mm}
\begin{center}
{\Large \bf All fundamental fermions fit inside one $su(1|5)$ irreducible representation}\\[5mm]
{\bf N.I.~Stoilova}~\footnote{Permanent address:
Institute for Nuclear Research and Nuclear Energy, Boul.\ Tsarigradsko Chaussee 72,
1784 Sofia, Bulgaria} 
{\bf and \underline{J.\ Van der Jeugt}}\\[2mm]
Department of Applied Mathematics and Computer Science,
University of Ghent,\\
Krijgslaan 281-S9, B-9000 Gent, Belgium.
\end{center}

\begin{abstract}
The Lie superalgebra $su(1|5)$ has irreducible representations of
dimension~32, in which the 32 fundamental fermions of one generation
(leptons and quarks, of left and right chirality, and their antiparticles)
can be accommodated. 
The branching of these $su(1|5)$ representations with respect to its
subalgebra $su(3)\times su(2)\times u(1)$ reproduces precisely the 
classification of these fundamental fermions according to 
the gauge group $su(3)^c\times su(2)^w\times u(1)^w$ of the
Standard Model. 
Furthermore, a simple construction of the relevant representations
is given, and some consequences are discussed.
\end{abstract}

{\bf Corresponding author}: J.\ Van der Jeugt,
Department of Applied Mathematics and Computer Science,
University of Ghent, Krijgslaan 281-S9, B-9000 Gent, Belgium. E-mail: Joris.VanderJeugt@UGent.be.
Tel. + 32 9 2644812. Fax. +32 9 2644995.
\vskip 5mm

\newpage
\section{Introduction}

The success of the Standard Model~\cite{EW,S}, with gauge group (or Lie algebra)
$su(3)\times su(2)\times u(1) \equiv su(3)^c\times su(2)^w\times 
u(1)^w$~\footnote{In this paper, we work only
with Lie algebras and Lie superalgebras, and not their Lie groups. This is why
we use the notation of Lie algebras rather than groups.}, is beyond doubt.
The theory is also full of unexplained patterns, and contains many free parameters.
In unified theories (``grand unified theories'') one tries to find
a simpler pattern by trying to fit the data of the Standard Model into a
larger unity. The classical example is that of Georgi and Glashow~\cite{SU5},
who proposed a unified theory based upon the Lie algebra $su(5)$.
Even though the $su(5)$ model has many nice features, it unified the
$su(3)\times su(2)\times u(1)$ model only partly: for example, the
fundamental fermions did not appear in a single irreducible representation
(irrep) of $su(5)$. Other models were proposed, of which the ones
based upon the Lie algebras $so(10)$~\cite{SO10} and $E_6$~\cite{E6} 
are the best known~\cite{Ross}.
Such unified theories had a setback when the proton decay, predicted
in the $su(5)$ model, was not confirmed by experiments.
Since then, particle physics turned its attention to supersymmetry,
superstrings, M-theory, $\ldots$

All this time, a unification in terms of the Lie superalgebra
$su(1|5)$ has been given little attention~\cite{NS,DJ,BKZ}. In this paper we point
out some elegant features of such a $su(1|5)$ model. The Lie superalgebra
$su(1|5)$ contains $u(1)\times su(3)^c\times su(2)^w\times u(1)^w$
as a subalgebra. So essentially it contains the Lie algebra of the
Standard Model as a subalgebra; the first $u(1)$ can be considered
as providing a label distinguishing between the same irreps of 
$su(3)^c\times su(2)^w\times u(1)^w$.
One particularly nice feature is that {\em all fundamental fermions of a single 
generation} (so all leptons and quarks of left and right chirality
and their antiparticles) {\em fit inside one single irreducible representation (irrep) of $su(1|5)$}.
Moreover, this irrep contains nothing else. 

In Section~2 we describe the relevant irreps of $su(1|5)$, and show how
the fundamental fermions of the first generation (the other generations
are similar) fit inside this. Section~3 gives some mathematical details
of the branching from $su(1|5)$ to $u(1)\times su(3)^c\times su(2)^w\times u(1)^w$.
Furthermore, it gives an interesting description of the $su(1|5)$ irreps
considered here. Some final remarks are given in Section~4.

\section{Fundamental fermions in $su(1|5)$}

Let us start by a description of the left handed fundamental
fermions in the Standard Model, i.e.\ the way they are grouped
into irreps of $su(3)^c\times su(2)^w\times u(1)^w$ (see, for example,
\cite{Slansky} or the classical reference books~\cite{refbooks}).
This description is given in Table~1. Herein, an
irrep of $su(3)^c\times su(2)^w\times u(1)^w$ is labelled
by $(\lambda \mu; j ; Y)$: $(\lambda \mu)$ is the Dynkin label
of the $su(3)^c$ irrep [so $(00)$ stands for the 1-dimensional
irrep, and the 3-dimensional irreps $(10)$ or $(01)$ are often 
referred to as {\bf 3} and {\bf 3}$^*$]; $(j)$ is the Dynkin
label of $su(2)^w$ [(0) is the 1-dimensional irrep; (1) is the
2-dimensional irrep with isospins $+1/2$ and $-1/2$]; 
and finally $Y$ is the $u(1)^w$ label corresponding to hypercharge.
In this table, $I_3^w$ is the notation for weak isospin, $Y=Y^w$ for
hypercharge, and $Q=I_3^w+Y/2$ for electromagnetic charge.

The corresponding antiparticles with right chirality appear in the contragredient
representations of the ones given in this table, that is: $(01;1;-\frac{1}{3})$ for the
right handed up and down antiquarks $(\tilde u_R, \tilde d_R)$; $(10;0;\frac{4}{3})$ for 
the right handed up quark $u_R$; $(10; 0; -\frac{2}{3})$ for the right handed down quark $d_R$;
$(00; 1; +1)$ for the right handed positron and antineutrino $(\tilde e_r, \tilde\nu_R)$; 
$(00; 0; -2)$ for the right handed electron $e_R$; and $(00; 0; 0)$ for the right handed neutrino $\nu_R$
(though there is some doubt about whether right handed neutrinos exist).

Let us now consider a particular class of representations of the Lie superalgebra $su(1|5)$. 
In general, irreps of $su(1|5)$ [or $sl(1|5)$] are labelled by five Kac-Dynkin labels~\cite{Kac}. 
Here, we only need the irreps with Kac-Dynkin labels $(p;0,0,0,0)$. 
Such a representation is {\em typical} if $p\not\in\{0,1,2,3,4\}$ and {\em atypical} otherwise~\cite{Kac,Kac2}.
If $p$ is real and $p>4$, then the irrep is unitary; 
the atypical irreps with $p\in\{0,1,2,3,4\}$ are also unitary~\cite{Gould}.
The typical irreps $(p;0,0,0,0)$ have dimension 32 and their decomposition or
branching to $u(1)\times su(3)^c\times su(2)^w\times u(1)^w$ is given by:
\begin{eqnarray}
(p;0,0,0,0) & \rightarrow & (\frac{5p}{4}; 00; 0; 0) + \nn\\
&& (\frac{5p}{4}-1; 10;0;-\frac{2}{3}) + (\frac{5p}{4}-1; 00;1;+1) + \nn\\
&& (\frac{5p}{4}-2; 01;0;-\frac{4}{3}) + (\frac{5p}{4}-2; 10;1;+\frac{1}{3}) 
  + (\frac{5p}{4}-2; 00;0;+2) + \nn\\
&& (\frac{5p}{4}-3; 10;0;+\frac{4}{3}) + (\frac{5p}{4}-3; 01;1;-\frac{1}{3}) 
  + (\frac{5p}{4}-3; 00;0;-2)+ \nn\\
&& (\frac{5p}{4}-4; 01;0;+\frac{2}{3}) + (\frac{5p}{4}-4; 00;1;-1) + \nn\\  
&& (\frac{5p}{4}-5; 00; 0; 0).
\label{p-branching}
\end{eqnarray}
Herein, the first label is just the new $u(1)$ value, and the remaining labels
are as previously introduced for $su(3)^c\times su(2)^w\times u(1)^w$.
It is easy to check that the dimensions in the right hand side of~(\ref{p-branching})
do indeed add up to 32. 

The atypical irreps have lower dimensions: 
\begin{eqnarray}
&& \dim(4;0,0,0,0)= 31,\ \dim(3;0,0,0,0)= 26,\ \nn\\
&& \dim(2;0,0,0,0)= 16,\ \dim(1;0,0,0,0)= 6,
\label{dim-atyp}
\end{eqnarray}
and of course the trivial irrep $(0;0,0,0,0)$ has dimension 1.
The decompositions of the atypical irreps~(\ref{dim-atyp}) 
to $u(1)\times su(3)^c\times su(2)^w\times u(1)^w$ can still
be obtained from~(\ref{p-branching}) by deleting those subalgebra irreps with nonpositive
$u(1)$ label. For example, the decomposition of $(4;0,0,0,0)$ is given by~(\ref{p-branching})
with $p=4$ but the last line deleted; the decomposition of $(3;0,0,0,0)$ is given by~(\ref{p-branching})
with $p=3$ but the last two lines deleted, etc.
These decompositions can be obtained from character formulas for typical
and singly atypical irreps~\cite{VHKT}.

It should now be clear that the 32-dimensional $su(1|5)$ irreps $(p;0,0,0,0)$
accommodate all fundamental fermions, and nothing more. This is once more 
summarized in Table~2.

Note that for $p=4$, one finds the same table but with the last line deleted,
so without the right handed neutrino (as some would prefer).

Observe that $su(1|5)$ representations were also used in~\cite{NS} to accommodate
fundamental fermions. In~\cite{NS}, the identification of basis states of the
representation with the fermions is different from the one given here
(in the sense that the $su(3)^c\times su(2)^w\times u(1)^w$
subalgebra structure is not maintained).
Moreover, in the construction of the present paper $p$ is in principle
arbitrary, and the representations considered are irreducible (see also next section).
The representation constructed in~\cite{NS} corresponds to $p=2$, and is a
32-dimensional {\em reducible} representation (the so-called Kac module~\cite{Kac,Kac2}).
The irreducible quotient of this representation is the 16-dimensional irrep
with labels $(2;0,0,0,0)$.

It should be added that the branching of $su(1|5)$ irreps with respect to $u(1)\times su(5)$
has already been described in~\cite{DJ}, and repeated in~\cite{BKZ}. These 
descriptions correspond to the branching given here in~(\ref{p-branching}).

In this paper, we shall give a particularly simple description of the basis
states of these irreps, including their transformation under the Lie superalgebra
generators of $su(1|5)$.

\section{The Lie superalgebra $su(1|5)$ and its irreducible representations $(p;0,0,0,0)$}

For an introduction to the theory of (simple) Lie superalgebras and their notation, see
\cite{Kac,Kac2} or~\cite{Scheunert}. The algebra
$su(1|5)$ is a real form~\cite{Kac,Parker} of the Lie superalgebra $sl(1|5)$, which is closely related to
$gl(1|5)$. A convenient basis of $gl(1|5)$ is given by the Weyl generators 
$e_{ij}$, with $i,j\in\{0,1,\ldots,5\}$. The grading of $gl(1|5)$ is as follows: the
{\em even} elements are given by $e_{00}$ and $e_{ij}$ with $i,j\in\{1,\ldots,5\}$; the
{\em odd} elements are $e_{0i}$ and $e_{i0}$ ($i=1,\ldots,5$). The Lie superalgebra
bracket (which stands for a commutator or an anticommutator) is determined by
\begin{equation}
\lb e_{ij}, e_{kl} \rb = \delta_{jk} e_{il} - (-1)^{\deg(e_{ij}) \deg(e_{kl})}
\delta_{il} e_{kj}, \label{Weyl}
\end{equation}
where $\deg(e_{ij})$ is 0 (resp.\ 1) if $e_{ij}$ is even (resp.\ odd).
One can define $sl(1|5)$ as the (super)commutator algebra of $gl(1|5)$;
its basis consists of all elements $e_{ij}$ ($i\ne j$) and
the Cartan elements $e_{00}+e_{ii}$ ($i=1,\ldots,5$). All such elements have
supertrace 0. The form
of $sl(1|5)$ corresponding to $su(1|5)$  satisfies
\begin{equation}
e_{ij}^*=e_{ji}.
\label{form}
\end{equation}

The representations $(p;0,0,0,0)$ have been studied extensively in~\cite{Palev80}.
Here we summarize some properties (see also~\cite{PSV}). The basis vectors or states of the
irrep $(p;0,0,0,0)$ are of the form
\beq
|p;\t\ra \equiv |p;\t_1,\t_2,\t_3,\t_4,\t_5 \rangle, \hbox{ with }
\t_i\in\{0,1\} \hbox{ and } |\t|=\sum_{i=1}^5 \t_i \leq p.
\label{states}
\eeq
So clearly, one finds back the dimension 32 for $p\geq 5$, and the condition
$|\t|\leq p$ is in agreement with the dimensions of the atypical irreps given
in~(\ref{dim-atyp}). 

The action of the diagonal generators on these states is given by:
\begin{eqnarray}
&& e_{00} |p;\t\ra = (p-|\t|) |p;\t\ra, \label{e00}\\
&& e_{ii} |p;\t\ra = \t_i |p;\t\ra. \label{eii}
\end{eqnarray}
The action of the odd generators $e_{0i}$ and $e_{i0}$ is also simple:
\begin{eqnarray}
&& e_{0i} |p;\t\ra = \t_i (-1)^{\t_1+\cdots+\t_{i-1}}
\sqrt{p-|\t|+1}\; |p;\t_1,\ldots,\t_i-1,\ldots,\t_n\ra, \label{e0i}\\
&& e_{i0} |p;\t\ra = (1-\t_i) (-1)^{\t_1+\cdots+\t_{i-1}}
\sqrt{p-|\t|} \; |p;\t_1,\ldots,\t_i+1,\ldots,\t_n\ra. \label{ei0}
\end{eqnarray}
The action of the remaining even generators $e_{ij}$ on these basis
states follows from the above and $e_{ij}=\lb e_{i0}, e_{0j}\rb$
($1\leq i\ne j\leq 5$):
\begin{eqnarray}
e_{ij} |p;\theta\ra &=& \t_{j}(1-\t_{ i})
(-1)^{\t_{ i}+\ldots +\t_{j-1}}
  \;|p;\ldots, \t_i+1, \ldots , \t_j-1 ,\ldots \ra, \;\hbox{for } i<j; \label{eij}\\
e_{ij} |p;\theta\ra &=& -\t_{j}(1-\t_{ i})
(-1)^{\t_{ j}+\ldots +\t_{i-1}}
  \;|p;\ldots, \t_j-1, \ldots , \t_i+1 ,\ldots \ra, \;\hbox{for } i>j. \label{eji}
\end{eqnarray} 

Note that the vectors $|p;\theta\ra$ form an orthonormal basis for the
irrep $(p;0,0,0,0)$, and that -- in agreement with the form~(\ref{form}) --
the representatives of the generators satisfy $e_{ij}^\dagger=e_{ji}$ 
with respect to this inner product.

We shall now describe some relevant subalgebras of $su(1|5)$.
First of all, the even subalgebra of $su(1|5)$ is $u(1)\times su(5)$.
Herein, the $su(5)$ basis consists of all elements $e_{ij}$ ($i\ne j$) with
$1\leq i,j \leq 5$ and the diagonal elements $e_{ii}-e_{i+1,i+1}$ ($1\leq i <5$).
The $u(1)$ generator should be an element with supertrace 0, commuting with
$su(5)$. It is unique up to a factor, and we choose:
\begin{equation}
X=\frac{1}{4} (5e_{00}+e_{11}+e_{22}+e_{33}+e_{44}+e_{55}).
\label{X}
\end{equation}
It is clear from the above actions (\ref{e00})-(\ref{eii}) that
\begin{equation}
X |p;\t\ra = (\frac{5}{4}p - |\t| ) |p;\t\ra.
\end{equation}
This yields the values of $u(1)$ in (\ref{p-branching}) or in Table~2.

Next, we consider the usual subalgebra $su(3)^c\times su(2)^w\times u(1)^w$ of $su(5)$.
In the current case, the generators of $su(3)^c$ are given by
$e_{ij}, e_{ji}$ ($1\leq i<j\leq 3$), $e_{11}-e_{22}$ and $e_{22}-e_{33}$.
Those of $su(2)^w$ by $e_{45}$, $e_{54}$ and $e_{44}-e_{55}$. In fact, the
usual weak isospin generator corresponds to $I^w_3=\frac{1}{2}(e_{44}-e_{55})$.
Finally, the $u(1)^w$ generator should be traceless and commuting with 
$su(3)^c\times su(2)^w$; it is again unique up to a factor, and in order to find
back the common unities one takes:
\begin{equation}
Y=Y^w= -\frac{2}{3} (e_{11}+e_{22}+e_{33}) + (e_{44}+e_{55}).
\label{Y}
\end{equation}

It is now easy to check the action of all these generators on the states
$|p;\t\ra$, and to associate a fundamental fermion with each of these
basis vectors. For example, $|p;00000\ra$ corresponds to the left handed
antineutrino $\tilde\nu_L$. The three basis vectors
\[
|p;10000\ra, \ |p;01000\ra, \ |p;00100\ra
\]
are the basis vectors of the $u(1)\times su(3)^c\times su(2)^w\times u(1)^w$ 
irrep $(\frac{5p}{4}-1; 10; 0; -\frac{2}{3})$, so they correspond to the
three right handed down quarks (one of each color). The two basis vectors
\[
|p;00010\ra, \ |p;00001\ra, 
\]
are the vectors of the $u(1)\times su(3)^c\times su(2)^w\times u(1)^w$ 
irrep $(\frac{5p}{4}-1; 00; 1; +1)$, and so they correspond to the
right handed positron $\tilde e_R$ and antineutrino $\tilde\nu_r$.
Continuing with this leads to Table~3 and a complete correspondence between
the basis vectors $|p;\t\ra$ and the fundamental fermions.

As already observed in~\cite{DJ,BKZ}, the adjoint representation of $su(1|5)$
can be associated with the usual 24 gauge bosons of $su(5)$, plus the
complex 5 of Higgs mesons (5+5) and an extra neutral vector boson. 
It follows from~(\ref{e00})-(\ref{eji}) that the action
of the 24 $su(5)$ generators on the basis states of $(p;0,0,0,0)$
is $p$-independent; however, the action of the remaining 11 $su(1|5)$
generators does depend upon~$p$. This opens the possibility that three
generations of fundamental fermions could be associated with three 
different $p$-values.

\section{Comments and conclusions}

In this paper we have presented a group theoretical framework for
unified model building in elementary particle physics. 
Elementary particles is not our field of specialization, so we
leave it to the specialists to consider this representation theoretic picture
as a basis for real models. 
It is only after such considerations that the proposed $su(1|5)$ structure
can be regarded as an interesting part of physics, or whether it is just 
a mathematical coincidence.

Let us nevertheless point out some peculiar properties of the present work.
First of all, $su(1|5)$ has $su(5)$ as a subalgebra, so the Georgi-Glashow
unification~\cite{SU5} is automatically built in. It is not clear to us,
however, whether a model based upon $su(1|5)$ would have the same difficulties
as that based upon $su(5)$.

An interesting feature of $su(1|5)$ is that all 32 fundamental fermions are
accommodated in one and the same irrep $(p;0,0,0,0)$, where $p>4$ can be
chosen arbitrary. For $p=4$, the irrep has dimension 31, and the bottom line
of Table~2 or Table~3 should be left out. In other words: the
right handed neutrino is then deleted from the set.

The Lie superalgebras $su(m|n)$ have the peculiar property that they possess
non-equivalent irreps with the same weight structure (up to a $u(1)$-shift),
a property not holding for (simple) Lie groups or Lie algebras.
For example, for two distinct values $p_1$ and $p_2$ greater than 4, the
irreps $(p_1;0,0,0,0)$ and $(p_2;0,0,0,0)$ are non-equivalent representations
of $su(1|5)$. But both have the same dimension (namely 32), and essentially the
same weight structure (up to a shift in the first $u(1)$ value). In other words,
the branching to the subalgebra $su(3)^c\times su(2)^w\times u(1)^w$ is the
same for both irreps. Such a feature is not possible for Lie algebras.
This property opens the possibility that the three known generations of
fundamental fermions could be associated with three non-equivalent
$su(1|5)$ irreps of the type $(p;0,0,0,0)$, all three having the 
same $su(3)^c\times su(2)^w\times u(1)^w$ subirreps.

As a final speculation, let us mention that a new kind of unification also
raises the question of a new kind of substructure.
In the case of $su(1|5)$, this could be offered by the quasi-fermi operators~\cite{PSV},
which behave similar to Fermi operators, and can be used as
``creation and annihilation operators'' precisely for the construction of
the irreps with labels $(p;0,0,0,0)$.

\section*{Acknowledgements}

N.I.\ Stoilova wishes to acknowledge Ghent University for a visitors grant.

\newpage
\section*{Tables}
\renewcommand{\arraystretch}{1.2}
\begin{table}[htb]
\caption{List of left handed fundamental fermions.}
\begin{center}
\begin{tabular}{|l|r|c|c|c|c|c|c|}
\hline
                          irrep & dim & fermions & symbol & $su(3)^c$ & $I_3^w$ & $Y^w$ & $Q$ \\
 labels                         &     &          &        & irrep     &         &       &     \\
\hline
$(10; 1; +\frac{1}{3})$ & 6 & up and down & $u_L$ & $3$ & $+1/2$ & $+1/3$ & $+2/3$ \\
\cline{4-8}
 & &  quarks & $d_L$ & $3$ & $-1/2$ & $+1/3$ & $-1/3$ \\
\hline
$(01; 0; -\frac{4}{3})$ & 3 & up antiquarks & $\tilde u_L$ & $3^*$ & $0$ & $-4/3$ & $-2/3$ \\ 
\hline
$(01; 0; +\frac{2}{3})$ & 3 & down antiquarks & $\tilde d_L$ & $3^*$ & $0$ & $+2/3$ & $+1/3$ \\ 
\hline
$(00; 1; -1)$ & 2 & electron and  & $e_L$ & $1$ & $-1/2$ & $-1$ & $-1$ \\ 
\cline{4-8}
 & & neutrino & $\nu_L$ & $1$ & $+1/2$ & $-1$ & $0$ \\ 
\hline
$(00; 0; +2)$ & 1 & positron  & $\tilde e_L$ & $1$ & $0$ & $+2$ & $+1$ \\ 
\hline
$(00; 0; 0)$ & 1 & antineutrino  & $\tilde \nu_L$ & $1$ & $0$ & $0$ & $0$ \\ 
\hline 
\end{tabular}
\end{center}
\end{table}

\begin{table}[htb]
\caption{All left and right handed fundamental fermions in the irrep $(p;0,0,0,0)$ of $su(1|5)$.}
\begin{center}
\begin{tabular}{|c|c|c|}
\hline
subalgebra irrep & dim & fermions \\
\hline
$(\frac{5p}{4}; 00; 0; 0)$  & 1 & $\tilde\nu_L$ \\
$(\frac{5p}{4}-1; 10; 0; -\frac{2}{3})$  & 3 & $d_R$ \\
$(\frac{5p}{4}-1; 00; 1; +1)$  & 2 & $(\tilde e_R, \tilde\nu_R)$ \\
$(\frac{5p}{4}-2; 01; 0; -\frac{4}{3})$  & 3 & $\tilde u_L$ \\
$(\frac{5p}{4}-2; 10; 1; +\frac{1}{3})$  & 6 & $(u_L, d_L)$ \\
$(\frac{5p}{4}-2; 00; 0; +2)$  & 1 & $\tilde e_L$ \\
$(\frac{5p}{4}-3; 00; 0; -2)$  & 1 & $e_R$ \\
$(\frac{5p}{4}-3; 01; 1; -\frac{1}{3})$  & 6 & $(\tilde u_R, \tilde d_R)$ \\
$(\frac{5p}{4}-3; 10; 0; +\frac{4}{3})$  & 3 & $u_R$ \\
$(\frac{5p}{4}-4; 01; 0; \frac{2}{3})$  & 3 & $\tilde d_L$ \\
$(\frac{5p}{4}-4; 00; 1; -1)$  & 2 & $(e_L, \nu_L)$ \\
$(\frac{5p}{4}-5; 00; 0; 0)$  & 1 & $\nu_R$ \\
\hline
\end{tabular}
\end{center}
\end{table}

\begin{table}[htb]
\caption{Association of fundamental fermions with basis vectors $|p;\t\ra$.}
\begin{center}
\begin{tabular}{|l|l|l|}
\hline
basis vectors $|p;\t\ra$     & subalgebra  & fermions \\
  & representation  &  \\ 
\hline \hline
$|p;00000\ra$ & $(\frac{5p}{4};00;0;0)$ & $\tilde{\nu}_L$ \\ 
\hline
$|p;10000\ra,\ |p;01000\ra, |p;00100\ra$ & $(\frac{5p}{4}-1;10;0;
-\frac{2}{3})$ & $d_R$ \\ 
\hline
$|p;00010\ra,\ |p;00001\ra $ & $(\frac{5p}{4}-1;00;1;+1$ & 
$\tilde{e}_R, \tilde{\nu}_R$ \\ 
\hline
$|p;11000\ra,\ |p;10100\ra, |p;01100\ra$ & $(\frac{5p}{4}-2;01;0;
-\frac{4}{3})$ & $\tilde{u}_L$ \\ 
\hline
$|p;10010\ra,\ |p;01010\ra, |p;00110\ra$ & $(\frac{5p}{4}-2;10;1;
+\frac{1}{3})$ & $u_L$ \\ 
$|p;10001\ra,\ |p;01001\ra, |p;00101\ra$ &  & $d_L$ \\ 
\hline
$|p;00011\ra$ & $(\frac{5p}{4}-2;00;0;+2)$ & $\tilde{e}_L$ \\ 
\hline
$|p;11100\ra$ & $(\frac{5p}{4}-3;00;0;-2)$ & $e_R$ \\ 
\hline
$|p;11010\ra,\ |p;10110\ra, |p;01110\ra$ & $(\frac{5p}{4}-3;01;1;
-\frac{1}{3})$ & $\tilde{d}_R$ \\ 
$|p;11001\ra,\ |p;10101\ra, |p;01101\ra$ &  & $\tilde{u}_R$ \\ 
\hline
$|p;10011\ra,\ |p;01011\ra, |p;00111\ra$ & $(\frac{5p}{4}-3;10;0;
+\frac{4}{3})$ & $u_R$ \\ 
\hline
$|p;11011\ra,\ |p;10111\ra, |p;01111\ra$ & $(\frac{5p}{4}-4;01;0;
+\frac{2}{3})$ & $\tilde{d}_L$ \\ 
\hline
$|p;11110\ra,\ |p;11101\ra $ & $(\frac{5p}{4}-4;00;1;
-1)$ & $\nu_L, e_L$ \\ 
\hline
$|p;11111\ra$ & $(\frac{5p}{4}-5;00;0;0)$ & $\nu_R$ \\ 
\hline
\end{tabular}
\end{center}
\end{table}

\end{document}